\documentclass[aps,prl,showpacs,twocolumn,superscriptaddress,groupedaddress]{revtex4}  % for review and submission

\usepackage{graphicx}
\usepackage{epsfig}

\usepackage{amssymb}
\usepackage{amsmath}

\usepackage{bm}        % for math
\usepackage{amssymb}   % for math
\usepackage{slashed}

\usepackage{lineno}
\usepackage{subfigure}

\usepackage{xspace}

\def\Dzero{D$0$\xspace}

\def\bc {\begin{center}}
\def\ec {\end{center}}
\def\beq {\begin{equation}}
\def\eeq {\end{equation}}

\def\NN  {\ensuremath{{{\sl N\kern -0.1em N}}\xspace}}

\def \tanb  {\ensuremath{\mathrm{\tan \beta}}\xspace}
\def \taum {\ensuremath{\mu}\xspace}
\def \tauh {\ensuremath{\tau_h}\xspace}
\def \b        {\ensuremath{b}\xspace}
\def \invfb       {\ensuremath{\mathrm{fb}^{-1}}}

\def \zmm {\ensuremath{Z/\gamma^*\to\mu^+\mu^-}\xspace}

\def\pt {\ensuremath{p_T}\xspace}
\def \pttau {\ensuremath{ p_T^{\tauh} }\xspace}
\def \ptmu {\ensuremath{ p_T^{\taum}}\xspace}
\def\pttrk{\ensuremath{p_T^{\mathrm{trk}} }\xspace}

\def \vpttau {\ensuremath{  \vec{p}_T^{\ \tauh}    }\xspace}
\def \vptmu {\ensuremath{  \vec{p}_T^{\ \taum}   }\xspace}

\def\met    {\ensuremath{\slashed{E}_T}\xspace}
\def\vmet    {\ensuremath{\vec{\slashed{E}}_T}\xspace}

\def\mD {\ensuremath{\mathcal{D}}\xspace}

\def\Dqcd {\ensuremath{\mD_{\text{MJ}}}\xspace}
\def\Dtt {\ensuremath{\mD_{\t\tbar}}\xspace}
\def\Df {\ensuremath{\mD_{f}}\xspace}

\def\RMJ{\ensuremath{R_{\text{iso}/\overline{\text{iso}}}}\xspace}

\def\mhat{\ensuremath{M_{\text{hat}}}\xspace}

\def\ggh{\ensuremath{gg\phi}\xspace}
\def\bbh{\ensuremath{\b\b\phi}\xspace}
\def\gbhb{\ensuremath{bgb\phi}\xspace}
\def\bbb{\ensuremath{\b\b\b}\xspace}

\def\hbb{$\phi\rightarrow b\bar{b}$}
\def\hbb{\bbb}

\def\BR         {{\ensuremath{\cal B}\xspace}}

\def\proton      {\ensuremath{p}\xspace}
\def\antiproton  {\ensuremath{\overline p}\xspace}
\def\c     {\ensuremath{c}\xspace}
\def\cbar  {\ensuremath{\overline c}\xspace}

\def\b     {\ensuremath{b}\xspace}
\def\bbar  {\ensuremath{\overline b}\xspace}

\def\t     {\ensuremath{t}\xspace}
\def\tbar  {\ensuremath{\overline t}\xspace}
\def\tbar  {\ensuremath{\overline t}\xspace}
\def\ttbar {\ensuremath{t\overline t}\xspace}

\def\piz   {\ensuremath{\pi^0}\xspace}

\newcommand{\tev}{\ensuremath{\mathrm{\,Te\kern -0.1em V}}\xspace}
\newcommand{\gev}{\ensuremath{\mathrm{\,Ge\kern -0.1em V}}\xspace}
\newcommand{\mev}{\ensuremath{\mathrm{\,Me\kern -0.1em V}}\xspace}
\newcommand{\kev}{\ensuremath{\mathrm{\,ke\kern -0.1em V}}\xspace}
\newcommand{\ev}{\ensuremath{\mathrm{\,e\kern -0.1em V}}\xspace}
\newcommand{\gevc}{\ensuremath{{\mathrm{\,Ge\kern -0.1em V\!/}c}}\xspace}
\newcommand{\mevc}{\ensuremath{{\mathrm{\,Me\kern -0.1em V\!/}c}}\xspace}
\newcommand{\gevcc}{\ensuremath{{\mathrm{\,Ge\kern -0.1em V\!/}c^2}}\xspace}
\newcommand{\mevcc}{\ensuremath{{\mathrm{\,Me\kern -0.1em V\!/}c^2}}\xspace}

\def\cm   {\ensuremath{{\rm \,cm}}\xspace}

\newcommand{\jprlBase}       {Phys.\ Rev.\ Lett.\xspace}
\newcommand{\jprBase}        {Phys.\ Rev.\xspace}
\newcommand{\jplBase}        {Phys.\ Lett.\xspace}

\newcommand{\nimBaseD}       {Nucl.\ Instrum.\ Methods\ Phys.\ Res.\xspace}

\newcommand{\nima}      [1]  {\nimBaseD~A~{\bf #1}}

\newcommand{\plb}       [1]  {\jplBase\ B~{\bf #1}}

\newcommand{\jprl}      [1]  {\jprlBase\ {\bf #1}}
\newcommand{\jprd}      [1]  {\jprBase\ D~{\bf #1}}
\begin{document}

%\begin{frontmatter}



\hspace{5.2in} \mbox{FERMILAB-PUB-11-674-E}
\title{Search for Higgs bosons of the minimal supersymmetric standard model in {\boldmath{$\proton\antiproton$}} collisions at {\boldmath$\sqrt{s}=1.96~\mathrm{\,Te\kern -0.1em V}$}}%with the D0 Experiment}

% remove these 3 lines before journal submittal.
%\centerline{author list dated 3 December 2011}
% end removal before journal submittal
%
\affiliation{Universidad de Buenos Aires, Buenos Aires, Argentina}
\affiliation{LAFEX, Centro Brasileiro de Pesquisas F{\'\i}sicas, Rio de Janeiro, Brazil}
\affiliation{Universidade do Estado do Rio de Janeiro, Rio de Janeiro, Brazil}
\affiliation{Universidade Federal do ABC, Santo Andr\'e, Brazil}
\affiliation{Instituto de F\'{\i}sica Te\'orica, Universidade Estadual Paulista, S\~ao Paulo, Brazil}
\affiliation{University of Science and Technology of China, Hefei, People's Republic of China}
\affiliation{Universidad de los Andes, Bogot\'{a}, Colombia}
\affiliation{Charles University, Faculty of Mathematics and Physics, Center for Particle Physics, Prague, Czech Republic}
\affiliation{Czech Technical University in Prague, Prague, Czech Republic}
\affiliation{Center for Particle Physics, Institute of Physics, Academy of Sciences of the Czech Republic, Prague, Czech Republic}
\affiliation{Universidad San Francisco de Quito, Quito, Ecuador}
\affiliation{LPC, Universit\'e Blaise Pascal, CNRS/IN2P3, Clermont, France}
\affiliation{LPSC, Universit\'e Joseph Fourier Grenoble 1, CNRS/IN2P3, Institut National Polytechnique de Grenoble, Grenoble, France}
\affiliation{CPPM, Aix-Marseille Universit\'e, CNRS/IN2P3, Marseille, France}
\affiliation{LAL, Universit\'e Paris-Sud, CNRS/IN2P3, Orsay, France}
\affiliation{LPNHE, Universit\'es Paris VI and VII, CNRS/IN2P3, Paris, France}
\affiliation{CEA, Irfu, SPP, Saclay, France}
\affiliation{IPHC, Universit\'e de Strasbourg, CNRS/IN2P3, Strasbourg, France}
\affiliation{IPNL, Universit\'e Lyon 1, CNRS/IN2P3, Villeurbanne, France and Universit\'e de Lyon, Lyon, France}
\affiliation{III. Physikalisches Institut A, RWTH Aachen University, Aachen, Germany}
\affiliation{Physikalisches Institut, Universit{\"a}t Freiburg, Freiburg, Germany}
\affiliation{II. Physikalisches Institut, Georg-August-Universit{\"a}t G\"ottingen, G\"ottingen, Germany}
\affiliation{Institut f{\"u}r Physik, Universit{\"a}t Mainz, Mainz, Germany}
\affiliation{Ludwig-Maximilians-Universit{\"a}t M{\"u}nchen, M{\"u}nchen, Germany}
\affiliation{Fachbereich Physik, Bergische Universit{\"a}t Wuppertal, Wuppertal, Germany}
\affiliation{Panjab University, Chandigarh, India}
\affiliation{Delhi University, Delhi, India}
\affiliation{Tata Institute of Fundamental Research, Mumbai, India}
\affiliation{University College Dublin, Dublin, Ireland}
\affiliation{Korea Detector Laboratory, Korea University, Seoul, Korea}
\affiliation{CINVESTAV, Mexico City, Mexico}
\affiliation{Nikhef, Science Park, Amsterdam, the Netherlands}
\affiliation{Radboud University Nijmegen, Nijmegen, the Netherlands and Nikhef, Science Park, Amsterdam, the Netherlands}
\affiliation{Joint Institute for Nuclear Research, Dubna, Russia}
\affiliation{Institute for Theoretical and Experimental Physics, Moscow, Russia}
\affiliation{Moscow State University, Moscow, Russia}
\affiliation{Institute for High Energy Physics, Protvino, Russia}
\affiliation{Petersburg Nuclear Physics Institute, St. Petersburg, Russia}
\affiliation{Instituci\'{o} Catalana de Recerca i Estudis Avan\c{c}ats (ICREA) and Institut de F\'{i}sica d'Altes Energies (IFAE), Barcelona, Spain}
\affiliation{Stockholm University, Stockholm and Uppsala University, Uppsala, Sweden}
\affiliation{Lancaster University, Lancaster LA1 4YB, United Kingdom}
\affiliation{Imperial College London, London SW7 2AZ, United Kingdom}
\affiliation{The University of Manchester, Manchester M13 9PL, United Kingdom}
\affiliation{University of Arizona, Tucson, Arizona 85721, USA}
\affiliation{University of California Riverside, Riverside, California 92521, USA}
\affiliation{Florida State University, Tallahassee, Florida 32306, USA}
\affiliation{Fermi National Accelerator Laboratory, Batavia, Illinois 60510, USA}
\affiliation{University of Illinois at Chicago, Chicago, Illinois 60607, USA}
\affiliation{Northern Illinois University, DeKalb, Illinois 60115, USA}
\affiliation{Northwestern University, Evanston, Illinois 60208, USA}
\affiliation{Indiana University, Bloomington, Indiana 47405, USA}
\affiliation{Purdue University Calumet, Hammond, Indiana 46323, USA}
\affiliation{University of Notre Dame, Notre Dame, Indiana 46556, USA}
\affiliation{Iowa State University, Ames, Iowa 50011, USA}
\affiliation{University of Kansas, Lawrence, Kansas 66045, USA}
\affiliation{Kansas State University, Manhattan, Kansas 66506, USA}
\affiliation{Louisiana Tech University, Ruston, Louisiana 71272, USA}
\affiliation{Boston University, Boston, Massachusetts 02215, USA}
\affiliation{Northeastern University, Boston, Massachusetts 02115, USA}
\affiliation{University of Michigan, Ann Arbor, Michigan 48109, USA}
\affiliation{Michigan State University, East Lansing, Michigan 48824, USA}
\affiliation{University of Mississippi, University, Mississippi 38677, USA}
\affiliation{University of Nebraska, Lincoln, Nebraska 68588, USA}
\affiliation{Rutgers University, Piscataway, New Jersey 08855, USA}
\affiliation{Princeton University, Princeton, New Jersey 08544, USA}
\affiliation{State University of New York, Buffalo, New York 14260, USA}
\affiliation{Columbia University, New York, New York 10027, USA}
\affiliation{University of Rochester, Rochester, New York 14627, USA}
\affiliation{State University of New York, Stony Brook, New York 11794, USA}
\affiliation{Brookhaven National Laboratory, Upton, New York 11973, USA}
\affiliation{Langston University, Langston, Oklahoma 73050, USA}
\affiliation{University of Oklahoma, Norman, Oklahoma 73019, USA}
\affiliation{Oklahoma State University, Stillwater, Oklahoma 74078, USA}
\affiliation{Brown University, Providence, Rhode Island 02912, USA}
\affiliation{University of Texas, Arlington, Texas 76019, USA}
\affiliation{Southern Methodist University, Dallas, Texas 75275, USA}
\affiliation{Rice University, Houston, Texas 77005, USA}
\affiliation{University of Virginia, Charlottesville, Virginia 22901, USA}
\affiliation{University of Washington, Seattle, Washington 98195, USA}
\author{V.M.~Abazov} \affiliation{Joint Institute for Nuclear Research, Dubna, Russia}
\author{B.~Abbott} \affiliation{University of Oklahoma, Norman, Oklahoma 73019, USA}
\author{B.S.~Acharya} \affiliation{Tata Institute of Fundamental Research, Mumbai, India}
\author{M.~Adams} \affiliation{University of Illinois at Chicago, Chicago, Illinois 60607, USA}
\author{T.~Adams} \affiliation{Florida State University, Tallahassee, Florida 32306, USA}
\author{G.D.~Alexeev} \affiliation{Joint Institute for Nuclear Research, Dubna, Russia}
\author{G.~Alkhazov} \affiliation{Petersburg Nuclear Physics Institute, St. Petersburg, Russia}
\author{A.~Alton$^{a}$} \affiliation{University of Michigan, Ann Arbor, Michigan 48109, USA}
\author{G.~Alverson} \affiliation{Northeastern University, Boston, Massachusetts 02115, USA}
\author{M.~Aoki} \affiliation{Fermi National Accelerator Laboratory, Batavia, Illinois 60510, USA}
% ...
%
% visitor_addresses.tex                        3 December 2011
%  available symbols are:
%  $\ast, \dag, \ddag, \S, \P, $\|$, $\ast\ast$, \dag\dag, \ddag\ddag ,\#
%
\collaboration{The D0 Collaboration\footnote{with visitors from
%{alton}
$^{a}$Augustana College, Sioux Falls, SD, USA,
%{burdin}
$^{b}$The University of Liverpool, Liverpool, UK,
%{falkowski}
%$^{?}$Laboratoire de Physique Theorique, Orsay, FR,
%{garcia-guerra}
$^{c}$UPIITA-IPN, Mexico City, Mexico,
%{haas,partridge}
$^{d}$SLAC, Menlo Park, CA, USA,
%{hesketh}
$^{e}$University College London, London, UK,
%{luna-garcia}
$^{f}$Centro de Investigacion en Computacion - IPN, Mexico City, Mexico,
%{podesta-lerma}
$^{g}$ECFM, Universidad Autonoma de Sinaloa, Culiac\'an, Mexico,
and 
%{weber}
$^{h}$Universit{\"a}t Bern, Bern, Switzerland.
%{grohsjean}
$^{i}$DESY, Hamburg, Germany,
%{hooper}
%$^{?}$Visitor from Bradley University, Peoria, IL, USA.
%{kozminski}
%$^{?}$}Visitor from Lewis University, Romeoville, IL, USA.
%{deceased}
$^{\ddag}$Deceased.
}} \noaffiliation
\vskip 0.25cm

%\date{\today}
\date{December 22, 2011}

\begin{abstract}
We report results from searches for neutral Higgs bosons produced in $\proton\antiproton$ collisions recorded by the \Dzero experiment at the Fermilab Tevatron Collider.  We study  the production of inclusive neutral Higgs boson in the $\tau\tau$ final state and in association with a $b$ quark in the $b\tau\tau$ and $\bbb$ final states. These results are combined to improve the sensitivity to the production of neutral Higgs bosons in the context of the minimal supersymmetric standard model (MSSM). 
The data are found to be consistent with expectation from background processes. Upper limits on MSSM Higgs boson production are set for Higgs boson masses ranging from $90$ to $300~\gev$. We exclude $\tanb>20-30$ for Higgs boson masses below $180~\gev$. These are the most stringent constraints on MSSM Higgs boson production in $\proton\antiproton$ collisions.
\end{abstract}

%\begin{keyword}


%\end{keyword}
%\end{frontmatter}
\maketitle

%\linenumbers
%%%%%%%%%%%%%%%%%%%%%%%%%%%%%%%%%%%%%%%%%%%%%%%%%%%%%%%%%%%%%
%%%%%%%%%%%%%%%%%%%%%%%%%%% Introduction %%%%%%%%%%%%%%%%%%%%%%%%%%%%
%%%%%%%%%%%%%%%%%%%%%%%%%%%%%%%%%%%%%%%%%%%%%%%%%%%%%%%%%%%%%
\section{Introduction}
\label{sec:intro}
In the minimal supersymmetric standard model (MSSM)~\cite{mssm},  the SU(2) symmetry is broken via two Higgs doublets; the first doublet couples to down-type fermions only while the second couples to up-type fermions. This leads to five physical Higgs bosons: two neutral {\it CP}-even bosons, $h$ and $H$, one neutral {\it CP}-odd boson $A$, and two charged bosons $H^\pm$. The neutral Higgs bosons are collectively denoted as $\phi$. 
At leading order the mass spectrum and the couplings of the Higgs bosons are determined by only two parameters, conventionally chosen to be  \tanb, the ratio of the two Higgs doublet vacuum expectation values, and $M_A$, the mass of the pseudoscalar Higgs boson. Radiative corrections introduce additional dependencies on other model parameters.  Although \tanb is a free parameter in the MSSM,  some indications suggest it should be large ($\tanb\gtrsim20$).  A value of  $\tanb\approx35$~\cite{cite:topbottom} would naturally explain the top to bottom quark mass ratio. The observed density of dark matter also points towards high \tanb values~\cite{cite:darkmatter}. 

\begin{figure*}
	\bc
	\subfigure[][]{\includegraphics[width=0.30\linewidth]{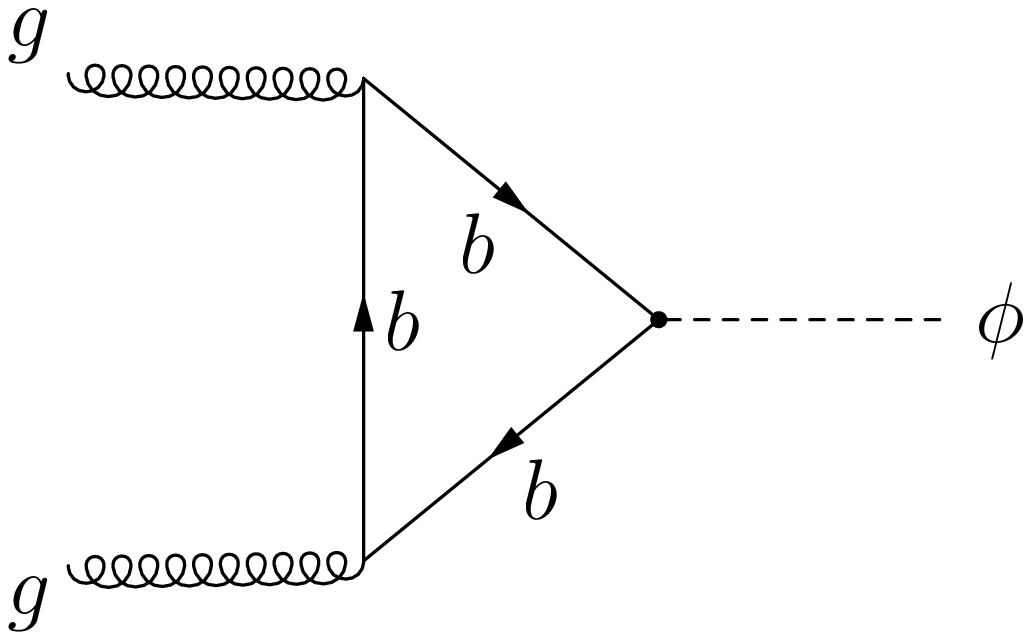}}\hspace{0.03\linewidth}
	\subfigure[][]{\includegraphics[width=0.30\linewidth]{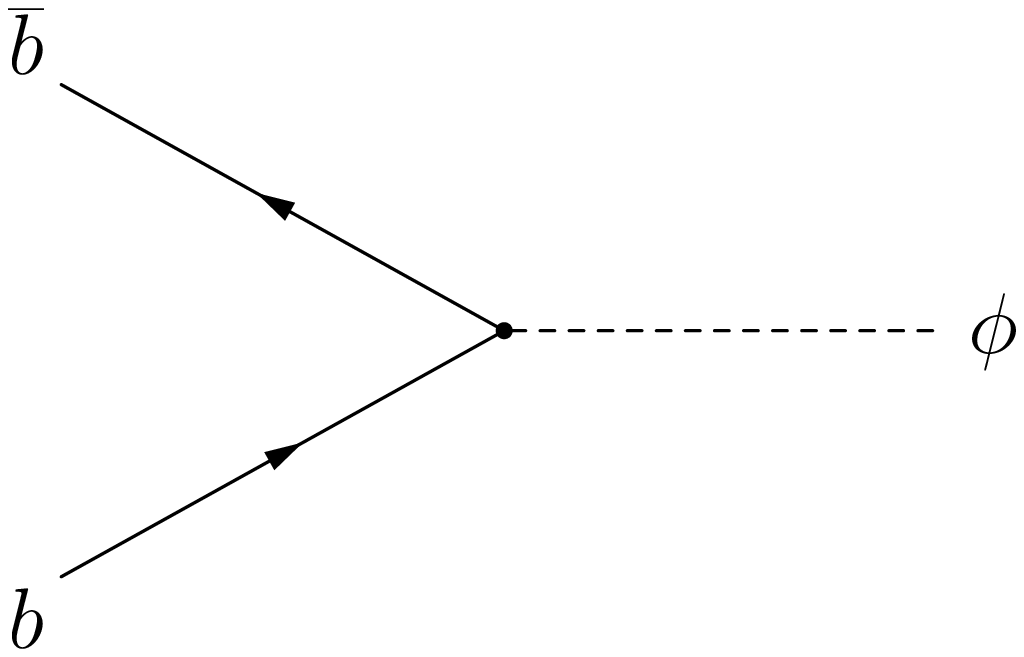}}\hspace{0.03\linewidth}
	\subfigure[][]{\includegraphics[width=0.30\linewidth]{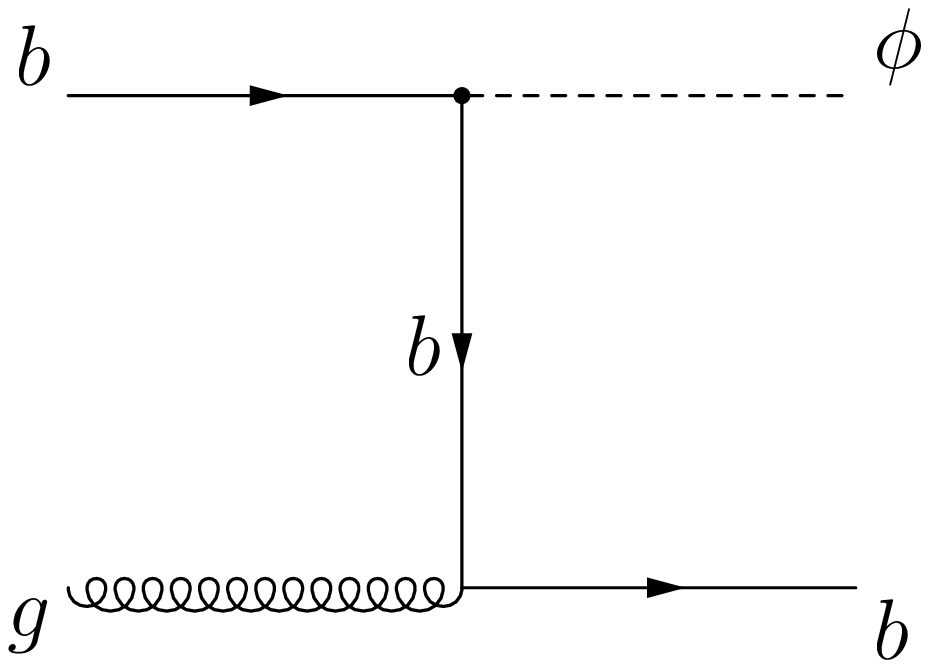}}
	
	\caption{Main Higgs boson production mechanisms in the MSSM in the 5-flavor scheme where \c and \b quarks are included in parton density functions. 
	The gluon fusion (a) and $b\bbar$ annihilation (b) processes dominate the inclusive production, while (c) is the dominant process for associated $b\phi$ production.}
	\label{fig:FeynGr}
	\ec
\end{figure*}

At large $\tanb$, one of the $CP$-even Higgs bosons ($h$ or $H$) is approximately degenerate in mass with the $A$ boson. In addition, they have similar couplings to fermions, which are enhanced (suppressed) by $\tanb$ compared to the standard model (SM) for down-type (up-type) fermions.
This enhancement has several consequences. First, the main decay modes become $\phi\to\b\bbar$ and $\phi\to\tau^+\tau^-$ with respective branching ratios $\BR(\phi\to\b\bbar)\approx90~\%$ and  $\BR(\phi\to\tau^+\tau^-)\approx10~\%$. Secondly, the main production processes  at a hadron collider involve $b$ quarks originating from the sea. Inclusive Higgs boson production is dominated by gluon fusion (\ggh) and $\b$$\bbar$ annihilation (\bbh), as shown in Fig.~\ref{fig:FeynGr}. The latter process may produce a $b$ quark in the acceptance of the detector in addition to the Higgs boson. This associated production $g\b\to\phi\b$ (\gbhb) is shown in Fig.~\ref{fig:FeynGr}c. In this case, the detection of the associated $b$ quark is a powerful experimental handle for reducing backgrounds.

MSSM Higgs boson masses below $93~\gev$ have been excluded by experiments at the CERN $e^+e^-$ Collider (LEP)~\cite{cite:LEP_exclu}. 
The CDF and \Dzero Collaborations have searched for MSSM neutral Higgs bosons decaying to tau pairs both inclusively~\cite{cite:CDF_tautau3,cite:D0_tautau3} and in association with a $b$ quark~\cite{cite:D0_btautau3}. The \Dzero Collaboration has also searched for $\b\phi\to\bbb$ production~\cite{cite:D0_bbb3}, which is challenging due to the high rate of multijet (MJ) production. Since these results have comparable sensitivities, combining them further enhances the potential reach. Recently, similar searches were performed at the LHC~\cite{cite:CMS_tautau,cite:ATLAS_tautau}. In this Letter, we present a combination of three searches performed by the \Dzero collaboration in the $\phi\to\tau\tau$, $b\phi\to\b\tau\tau$, and $b\phi\to\bbb$ final states. Since the inclusive and \gbhb production signal samples in the di-tau final states are not mutually exclusive, the \Dzero result presented in~\cite{cite:D0_tautau3} can not be directly combined with~\cite{cite:D0_btautau3}. Hence, we re-analyse here the inclusive $\phi\to\tau\tau$ production: we require that there are no $b$ jets, we extend the dataset to $7.3~\invfb$ of integrated luminosity, and we increase the trigger acceptance and refine the treatment of systematic uncertainties. The di-tau channels are restricted to final states where one $\tau$ lepton ($\tau_\mu$) decays via  $\tau\to\mu\nu_\mu\nu_\tau$ and the other ($\tauh$) decays hadronically. 
	
\section{Detector and object reconstruction}
The data analysed in the different studies presented here have been recorded by the \Dzero detector~\cite{run2det}. 
It has a central-tracking system, consisting of a silicon microstrip tracker and a central fiber tracker, both located within a 2~T superconducting solenoidal magnet, with designs optimised for tracking and vertexing at pseudorapidities~\cite{pseudorapidity} $|\eta|<3$ and $|\eta|<2.5$, respectively. A liquid-argon and uranium calorimeter has a  central section covering pseudorapidities  $|\eta|$ up to $\approx 1.1$, and two end calorimeters that extend coverage to $|\eta|\approx 4.2$, with all three housed in separate cryostats~\cite{run1det}. An outer muon system, at $|\eta|<2$,  consists of a layer of tracking detectors and scintillation trigger counters in front of 1.8~T toroids, followed by two similar layers after the toroids.

\begin{table}[h]
	\bc
	\caption{Searches combined in this Letter.}
	\label{tab:channels_summary}
	\begin{tabular}{l c c}
	\hline\hline
	\multicolumn{1}{c}{Final state}     & $\mathcal{L}$ (\invfb)   & Reference\\
	\hline
	$\phi\to\tau_\mu\tauh$  ($b$-jet veto)  & 7.3 & \\
	$\b\phi\to\b\tau_\mu\tauh$                     & 7.3 & \cite{cite:D0_btautau3} \\
	$\b\phi\to\b\bbar\b$                                 & 5.2 & \cite{cite:D0_bbb3} \\
	\hline\hline
	\end{tabular}
	\ec
\end{table}

%%% A word on triggers
The integrated luminosities ($\mathcal{L}$)~\cite{d0lumi} associated with each search are summarized in Table~\ref{tab:channels_summary}. Di-tau events were recorded using a mixture of single high-\pt muon, jet, tau, muon plus jet, and muon plus tau triggers.  The efficiency of this inclusive trigger condition is measured in a $Z\to\tau_\mu\tauh$ data sample with respect  to single muon triggers. We also verify this measurement in a sample of $Z(\to\tau_\mu\tauh)$+jets events. Depending on the kinematics and on the decay topology of the $\tauh$, the trigger efficiency ranges from $80\%$ to $95\%$. For the \bbb analysis, we employ triggers selecting events with at least three jets. Most of the \bbb data sample was  recorded with $b$-tagging requirements at the trigger level. The trigger efficiency for $m_\phi=150~\gev$ is approximately $60\%$ for events passing the analysis requirements.

Muons are reconstructed  from track segments in the muon system. They are matched to tracks in the inner tracking system. The timing of associated hits in the scintillators must be consistent with the beam crossing to veto cosmic muons. 

Hadronic tau decays  are characterised by narrow jets that are reconstructed using a jet cone algorithm with a radius of $0.3$~\cite{cone} in the calorimeter and by low track multiplicity~\cite{nntau}. We split the $\tauh$ candidates into three different categories that approximately correspond to one-prong $\tau$ decays with no \piz meson ($\tauh$ type 1), one-prong decay with \piz mesons ($\tauh$ type 2), and multi-prong decay ($\tauh$ type 3). In addition,  a neural-network-based $\tauh$ identification ($\NN_{\tau}$) has been trained to discriminate light parton jets ($u$, $d$, $s$ quarks or $\rm gluon$) from hadronic $\tau$ decays~\cite{nntau}.  
We select \tauh candidates requiring $\NN_\tau>0.9$ (0.95 for $\tauh$ type 3). This condition has an efficiency of approximately $65\%$ while rejecting $\sim$$99\%$ of quark/gluon jets. 

Jets are reconstructed from energy deposits in the calorimeter~\cite{cite:jetx} using the midpoint cone algorithm~\cite{cone} with a radius of~$0.5$. All jets are required to have at least two reconstructed tracks originating from the $\proton\antiproton$ interaction vertex matched within $\Delta R($track, jet-axis$)=\sqrt{(\Delta\eta)^2 +(\Delta\varphi)^2}<0.5$ (where $\varphi$ the azimuthal angle). To identify jets originating from $b$ quark decay, a neural network $b$-tagging algorithm ($\NN_b$)~\cite{nnbtag} has been developed. It uses lifetime-based information involving the track impact parameters and secondary vertices as inputs. 

The presence of neutrinos is inferred from the missing transverse energy, \met, which is reconstructed as the negative of the vector sum of the transverse energy of calorimeter cells with $|\eta|<3.2$, corrected for the energy scales of all reconstructed objects and  for muons.
\begin{figure*}[!t]
	\bc
	\includegraphics[width=0.35\linewidth]{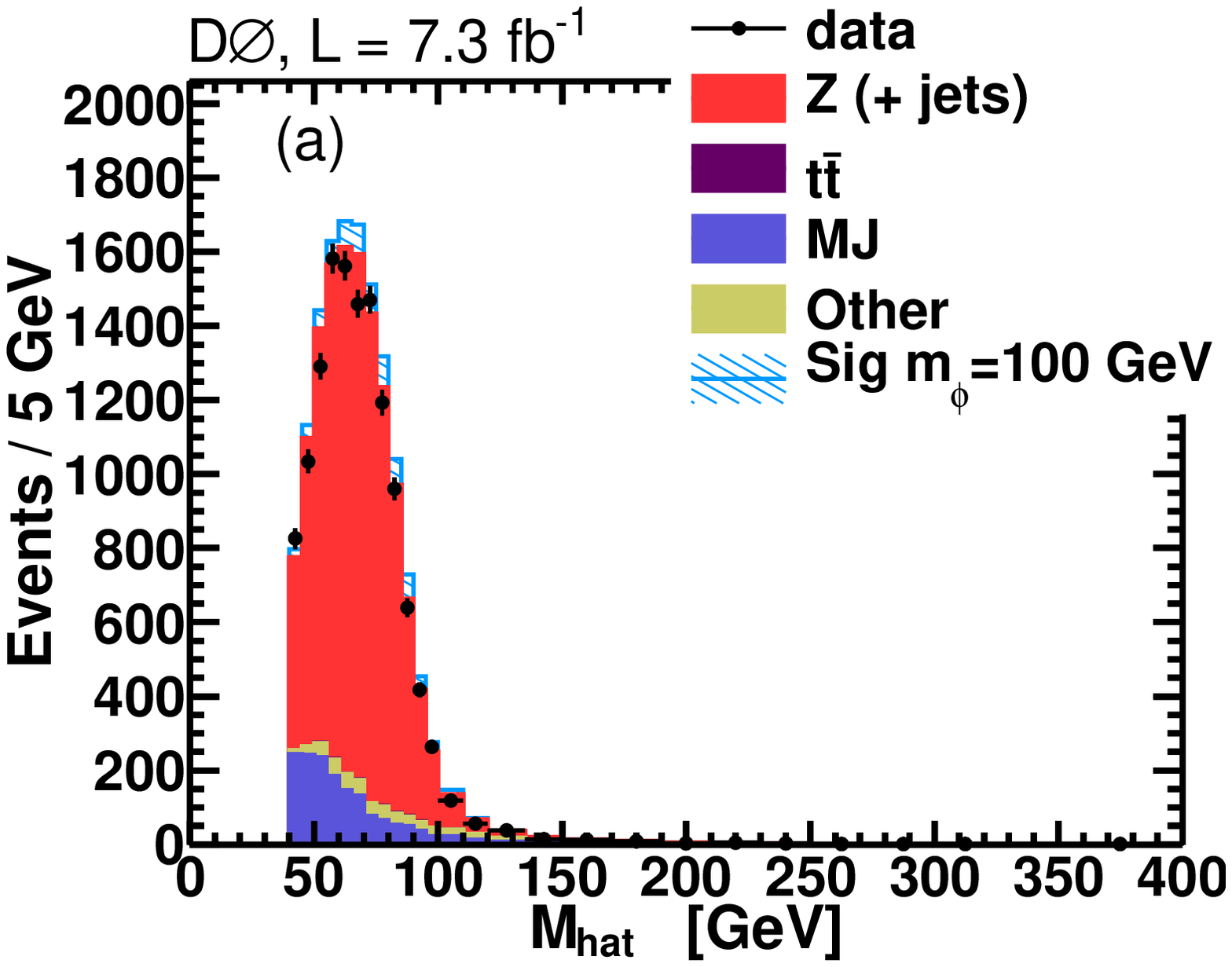}\hspace{0.06\linewidth}
	\includegraphics[width=0.35\linewidth]{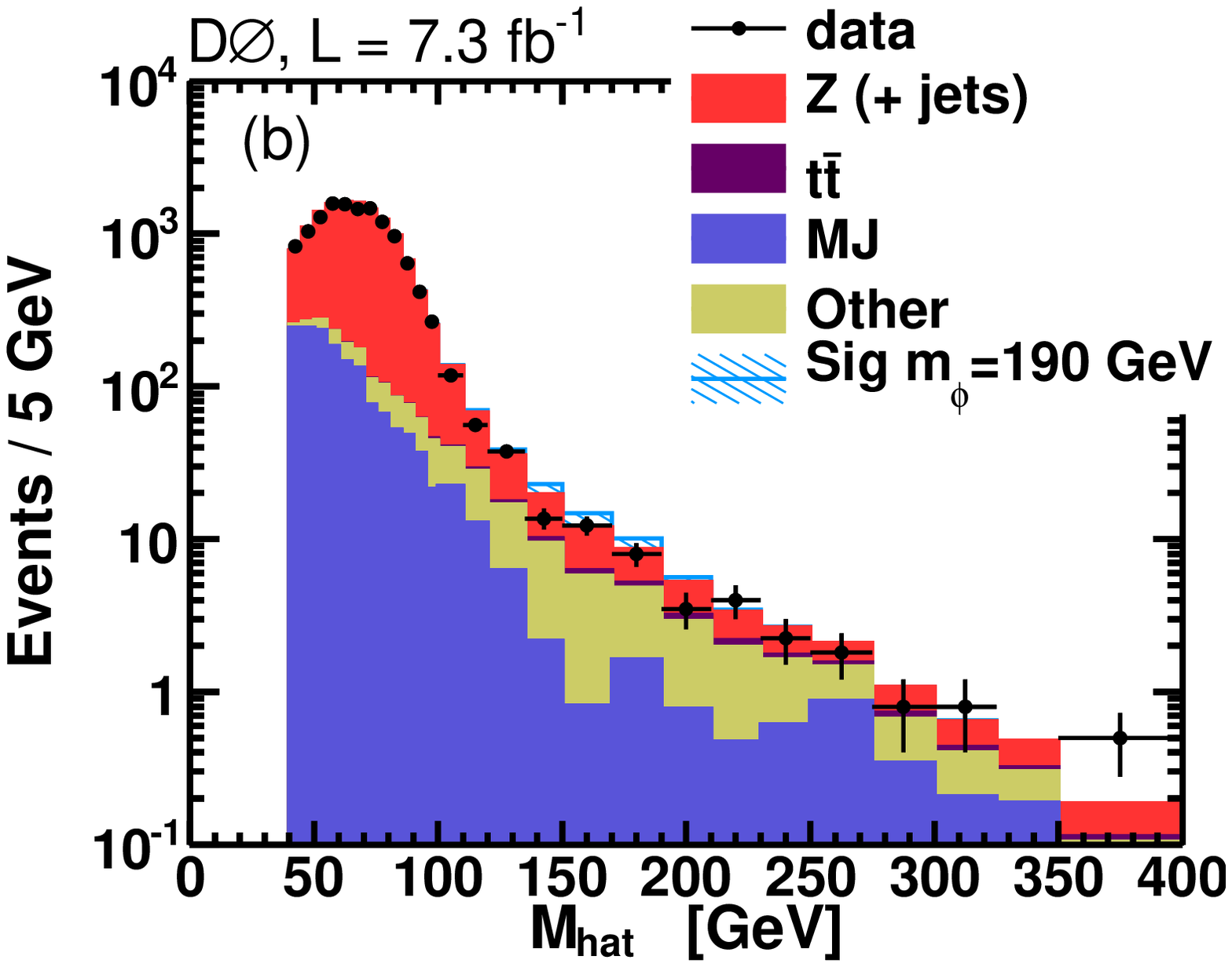}
	\includegraphics[width=0.35\linewidth]{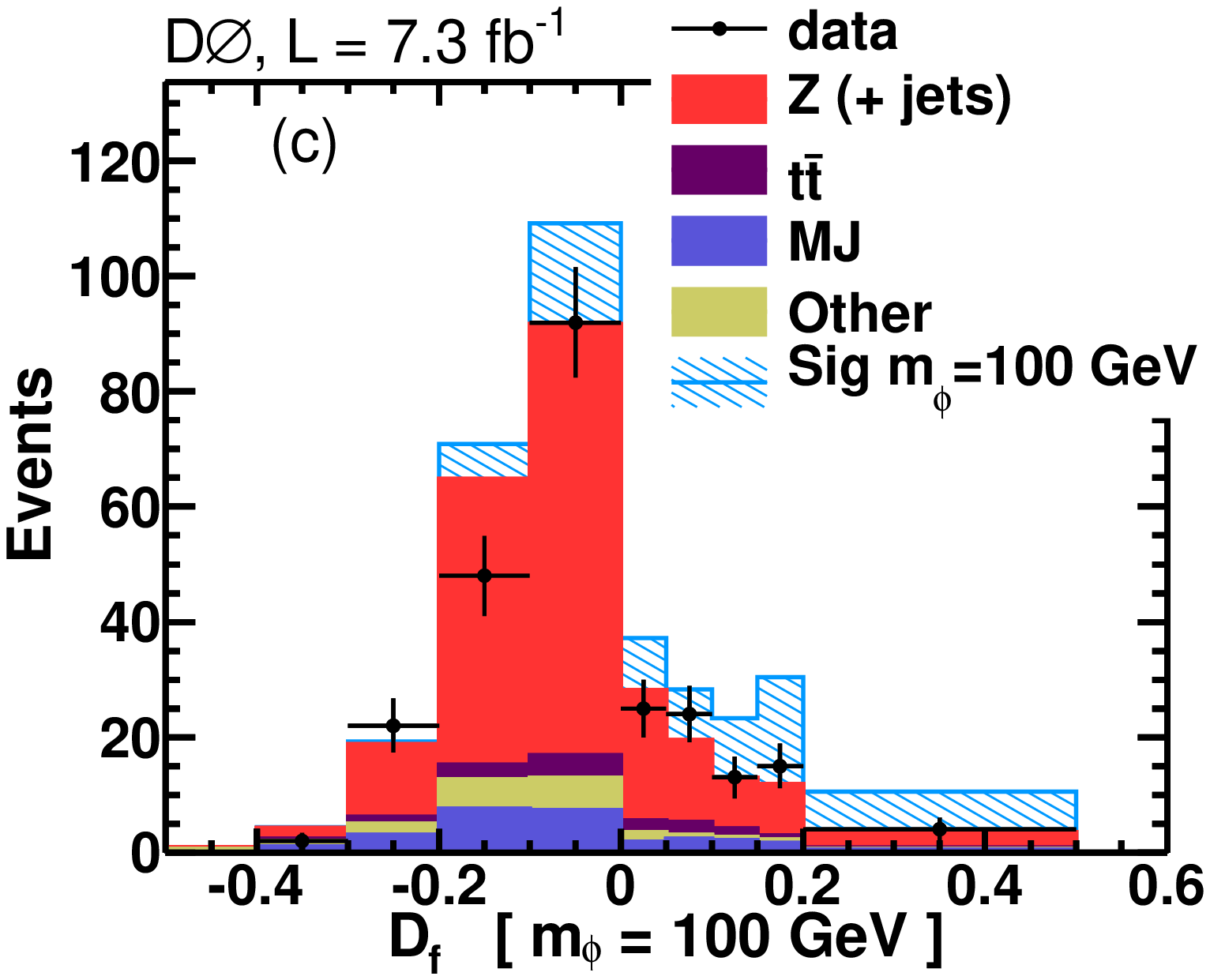}\hspace{0.06\linewidth}
	\includegraphics[width=0.35\linewidth]{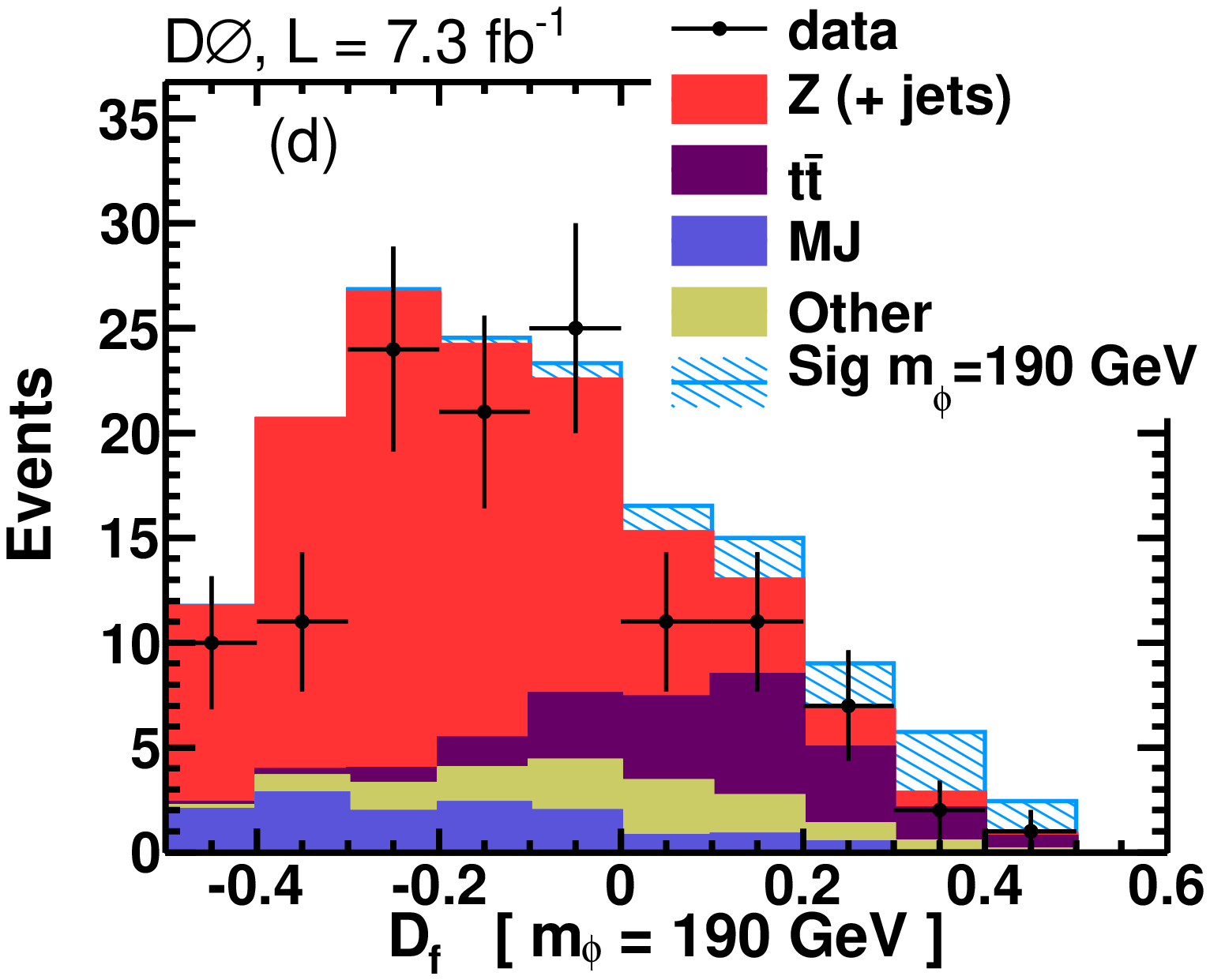}	
	\caption{ Distribution of $M_{hat}$ in the inclusive $\tau\tau$ sample on (a) linear and (b) logarithmic scale. 
		       (c) $\Df$ in the $b\tau\tau$ sample trained for $m_\phi=100~\gev$, and (d)  for $m_\phi=190~\gev$ adding the final requirements on $\Dqcd$ and $\Dtt$.  All \tauh types are combined.
		       The predicted signal is shown in the case of the $m_h^{\text{max}}$ scenario ($\mu=+200\gev$ and $\tan\beta=40$).}
	\label{fig:ditau_vars}
	\ec
\end{figure*}

%%%%% MC simulation
\section{Signal and background Monte Carlo simulation}

Signal samples are generated with the LO event generator  {\sc pythia}~\cite{pythia}. The inclusive production is simulated with the SM $\ggh$ process. We checked that the kinematic differences between $\bbh$ and \ggh do not have any impact on our final result. The associated production with a $b$-quark is generated with the SM $g\b\to\phi\b$ process.
The contributions to the $\b\phi$ cross section and event kinematics from next-to-leading order (NLO) diagrams are taken into account by using {\sc mcfm}~\cite{mcfm} to calculate correction factors for the {\sc pythia} generator as a function of the leading $b$ quark \pt and $\eta$ in the range $p_T^b>12~\gev$ and $|\eta^b|<5$.

In the final states with a tau pair, the dominant backgrounds are due to $Z\to\tau\tau$(+jets), diboson ($WW$, $WZ$ and $ZZ$), $W$+jets, \ttbar pair and MJ production, the latter being estimated from data. Diboson events are simulated with {\sc pythia} while the $Z$+jets, $W$+jets, and $\ttbar$ samples are generated using {\sc alpgen}~\cite{alpgen}.
In the \bbb channel, the dominant background is due to MJ production. We simulate MJ background events from the $\b\bbar j$, $\b\bbar jj$, $\c\cbar j$, $\c\cbar jj$, $\b\bbar\c\cbar$, and $\b\bbar\b\bbar$ processes, where $j$ denotes a light parton, with the {\sc alpgen} event generator. The small contribution from $\t\tbar$ production to the background is also simulated with {\sc alpgen}. The contribution from other processes, such as $Z+b\bar{b}$ and single top quark production, is negligible.

The {\sc alpgen} samples  are processed through {\sc pythia} for showering and hadronization. {\sc tauola}~\cite{tauola} is used to decay $\tau$ leptons and  {\sc evtgen}~\cite{evtgen}  to model $b$ hadron decays. All   samples are further processed through a detailed {\sc geant}~\cite{geant}-based simulation of the \Dzero detector. The output is then combined with data events recorded during random beam crossings to model the effects of detector noise and pile-up energy from multiple interactions and different beam crossings. Finally, the same reconstruction algorithms as for data are applied to the simulated events. Data control samples are used to correct the simulation for object identification efficiencies, energy scales and resolutions, trigger efficiencies, and the longitudinal $\proton\antiproton$ vertex distribution. Signal, \ttbar pair, and diboson yields are normalised to the product of their acceptance and detector efficiency (both determined from the simulation), their corresponding theoretical cross section and the luminosity.   

In the \bbb final state, the relative contribution of the different MJ backgrounds is determined from data; its overall normalisation is constrained by a fit done in the final limit-setting procedure which exploits  the dijet-mass shape differences between signal and background. In the di-tau channels, a dedicated treatment of the dominant $Z\to\tau\tau$ background has been developed to reduce its systematic uncertainties. The  simulation of the $Z$ boson kinematics is corrected by comparing a large sample of $Z\to\mu\mu$ events in data and in the simulation. We measure correction factors in each jet multiplicity bin as a function of the $\Phi^*$ quantity introduced in Ref.~\cite{phistar}, leading jet $\eta$, and leading $b$-tagged jet $\NN_b$. This affects both the normalisation and the kinematic distributions. For the $W$+jets background, the muon predominantly arises from the $W$ boson decay while the  $\tauh$ candidate is a misreconstructed jet.  
The $W$+jets simulation is normalised to data, for each jet multiplicity bin, using a $W(\to\mu\nu)+$jets data control sample. 

\begin{figure}[!tb]
	\bc
\includegraphics[width=0.85\linewidth,height=0.98\linewidth]{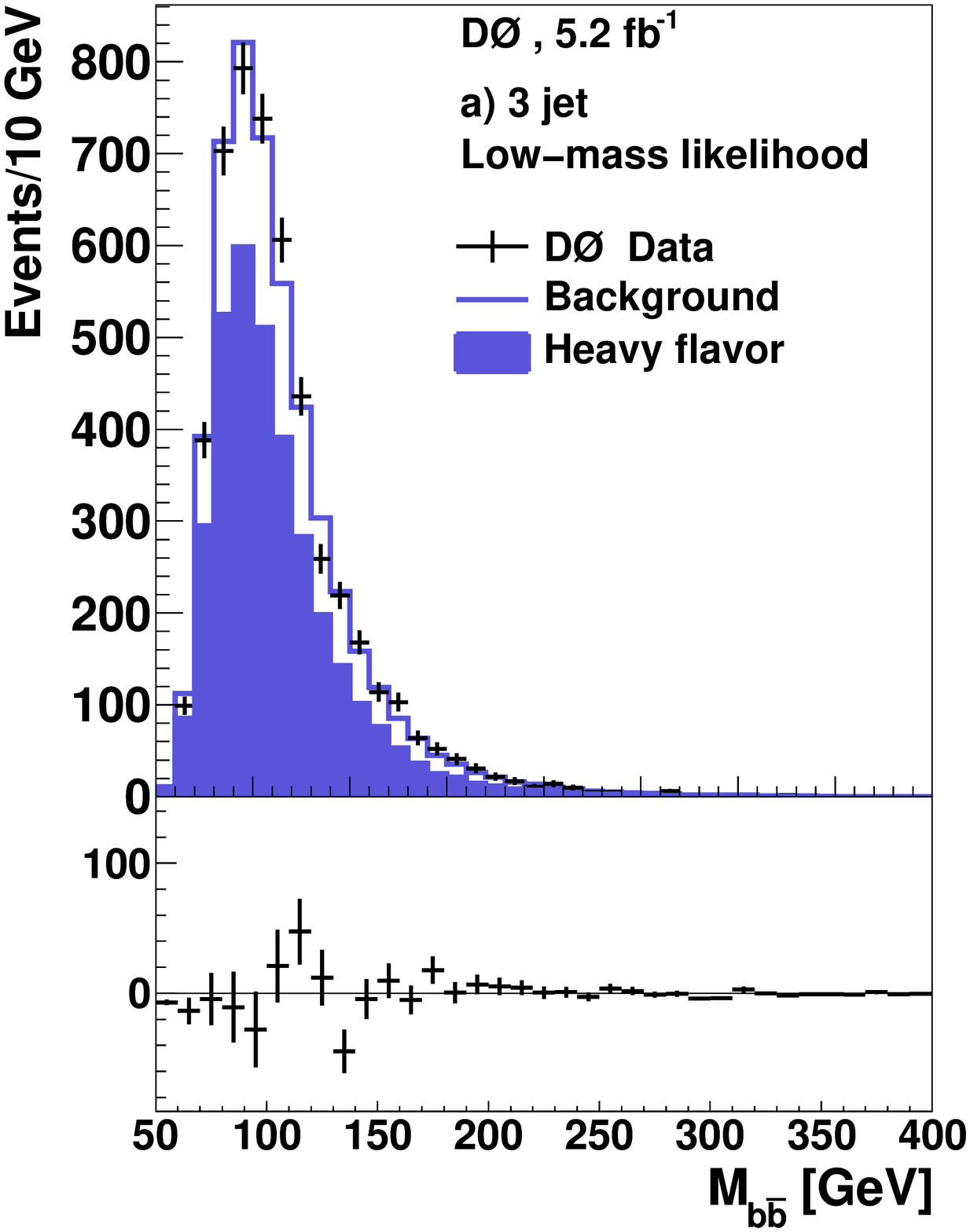}\\%  \hspace{0.06\linewidth}
\includegraphics[width=0.85\linewidth,height=0.98\linewidth]{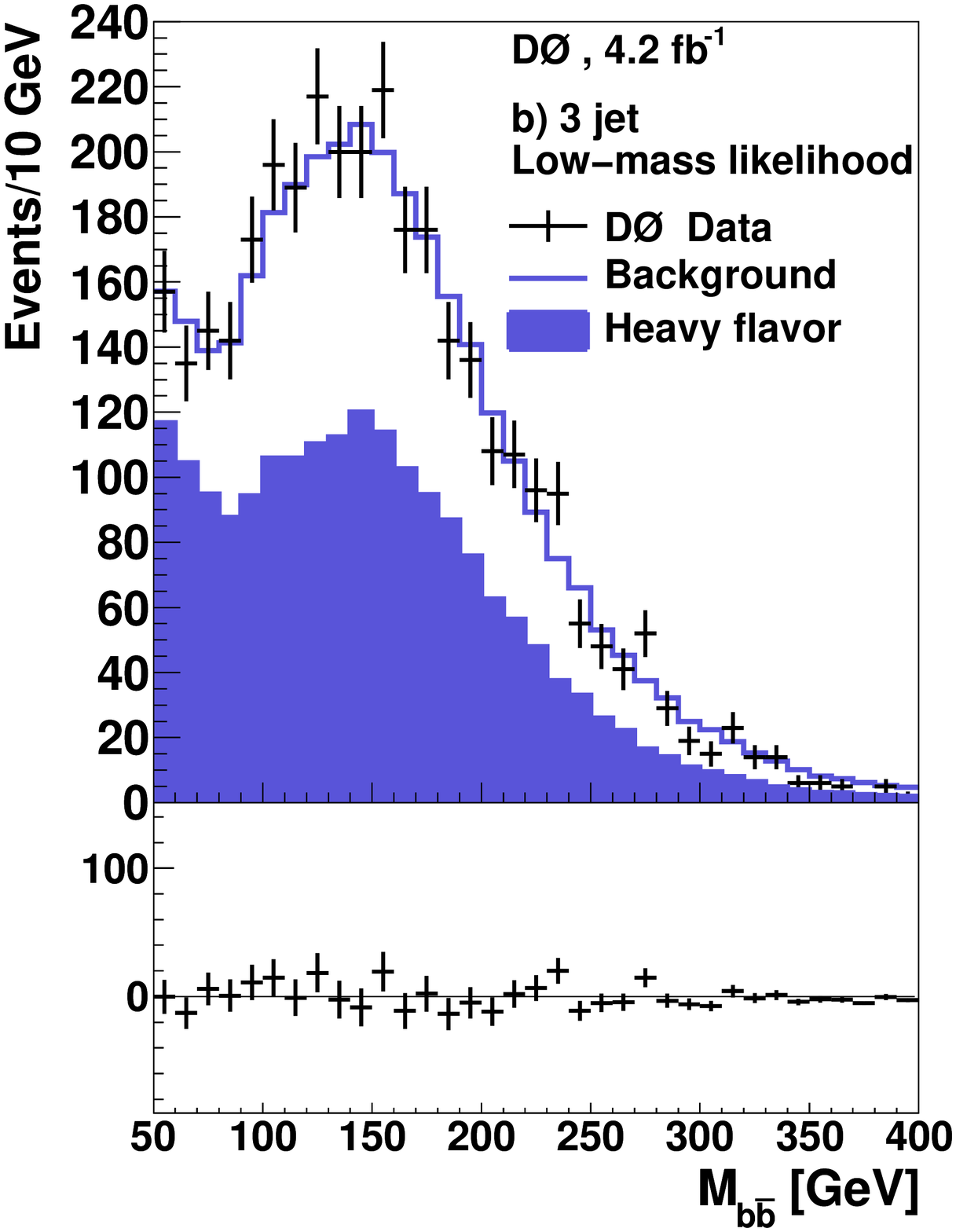} 
	\caption{Distributions of the dijet invariant mass, taken from Ref.~\cite{cite:D0_bbb3}, in the signal region defined by $\mathcal{D}_{\bbb}>0.65$ (a) and in a control region defined by $\mathcal{D}_{\bbb}<0.12$ (b) for the dominant \hbb\ 3-jet channel is shown. The line shows the background model, the solid histogram the component coming from \bbb, the points with error bars show the data. The background is normalized to the data yield for illustration purposes. The difference between data and the background model is shown at the bottom of each panel.}
\label{fig:inputshbb}
	\ec
\end{figure}

\section{Analysis strategy}

In this section, we describe the search strategy as well as the selection of the final signal samples. Further details of the $\b\tau\tau$ and $\bbb$ analyses can be found in Refs.~\cite{cite:D0_btautau3} and~\cite{cite:D0_bbb3}, respectively.

\subsection{Di-tau final states}

The $\tau\tau$ and  $\b\tau\tau$ searches follow a similar strategy. We first define a common selection by retaining events with one reconstructed $\proton\antiproton$ interaction vertex with at least three tracks, exactly one isolated muon and exactly one reconstructed \tauh. We require the muon to have a transverse momentum $\ptmu > 15~\gev$, $|\eta^{\taum}| <1.6$, and to be isolated in the calorimeter and in the central tracking system, \textit{i.e.}, $\Delta R(\taum,\text{jet})>0.5$ 
relative to any reconstructed jet. 
The  \tauh candidate must have a transverse momentum, as measured in the calorimeter with appropriate energy corrections, $\pttau>10~\gev$, $|\eta_{\tauh}| < 2.0$, $\Delta R(\tauh,\taum)>0.5$ relative to any muon, and $\tauh$ tracks must not be shared with any reconstructed muons in the event. 
The sum of the transverse momenta, \pttrk, of all tracks associated with the $\tauh$ candidate must satisfy $\pttrk>7/5/10~\gev$, respectively, for \tauh types 1/2/3. 
We require the distance along the beam axis between the \tauh and  the muon, at their point of closest approach to the $p\bar{p}$ interaction vertex, $\Delta z(\tauh,\taum) < 2~\cm$.  In addition, the \tauh and the muon must have an opposite electric charge (OS) and a transverse mass $M_T(\taum,\met)< 60~\gev$ (100~\gev for $\tauh$ type 2) where $M_T(\taum,\met)=\sqrt{2\cdot\ptmu\cdot\met\cdot\left[1- \cos \Delta \varphi(\taum,\met)\right]}$.

\begin{table}[t]
	\caption{Expected background yield, observed data yield,
	 and expected signal yields for the di-tau selections with their total systematic uncertainties. 
	 The signal yields are given for the $m_h^{\text{max}}$ scenario ($\mu=+200\gev$ and $\tan\beta=40$). } 
	\label{tab:ditau_yields}
	\begin{tabular}{l  @{\hspace{0.5cm}} r@{$\ \pm\ $}r @{\hspace{0.5cm}}  r@{$\ \pm\ $}r}
 	\hline\hline
	& \multicolumn{2}{c}{$\tau\tau$} & \multicolumn{2}{c}{$\b\tau\tau$} \\
	\hline
	$Z$(+jets)   & 11547 & 634 & 218 & 17\\	
	$t \bar{t}$    &        25 &      4 & 183 & 32\\
	MJ                &   1343 &  236 &   36 & 6 \\
	Other           &     560  &    25 &   40 & 2 \\
	\hline
	Total background            &  13474 & 684  & 476 & 40 \\
	Data                 &  \multicolumn{2}{c}{ 13344} &  \multicolumn{2}{c}{488}\\
	\hline
Signal $m_\phi=100~\gev$ & \multicolumn{2}{c}{1165} & \multicolumn{2}{c}{ 81 }\\
Signal $m_\phi=190~\gev$ & \multicolumn{2}{c}{    70} & \multicolumn{2}{c}{ 12}\\
\hline\hline
	\end{tabular}
\end{table}

\subsubsection*{Inclusive $\tau\tau$ selection}
For the inclusive $\tau\tau$ selection, we tighten the requirements on the \tauh transverse momentum to suppress the MJ background: $\pttau>12.5~\gev$ ($15~\gev$ for \tauh type 3) and $\pttrk>12.5/7/15~\gev$ respectively for \tauh type 1/2/3. We further reduce the $W$+jets background by requiring $M_T(\taum,\met)< 40~\gev$. 
We define $\mhat$, which represents the minimum center-of-mass energy consistent with the decay of a di-tau resonance, by
$$\mhat\equiv\sqrt{\left( E^{\taum\tauh} - p_{z}^{\taum\tauh}  + \met  \right)^2 - | \vpttau + \vptmu + \vmet |^2},$$ where $E^{\taum\tauh}$ is the energy of the $\taum\tauh$ system and  $p_z^{\taum\tauh}$ is its momentum component along the beam axis. We require $\mhat>40~\gev$ to suppress the MJ background. Finally, to prevent any overlap with the $\b\tau\tau$ sample, we select only events for which no jet has $\NN_\b>0.25$.

\subsubsection*{$\b\tau\tau$ selection}
The complementary sample with at least one $b$-tagged jet with $\NN_b>0.25$  constitutes the $\b\tau\tau$ sample. This $b$-tagged sample suffers from large $Z$+jets, \ttbar and MJ backgrounds. We build separate multivariate discriminants, $\Dqcd$ and $\Dtt$, to discriminate against the MJ and $\ttbar$ processes. We require $\Dqcd>0.1$ and $\Dtt>0.1$, then we combine $\NN_b$, $\Dqcd$, and $\Dtt$, to form a set  of final discriminating variables $\Df$ (one for each \tauh type and $m_\phi$) to be used in the limit-setting procedure. Further details can be found in Ref. ~\cite{cite:D0_btautau3}.

\subsubsection*{MJ background estimation}
In both di-tau channels, the MJ background is estimated from data control samples applying two different methods. 
The first is based on the small correlation between the electric charge of muon and \tauh in MJ events.  For each analysis, we select a data sample with identical criteria as the signal sample but with the two leptons having the same electric charge (SS). We subtract the residual contribution from other SM backgrounds from this MJ-dominated SS sample. We measure the ratio of the number of OS to SS events  to be $1.09\pm0.01$ and $1.07\pm0.01$, respectively, in the $\tau\tau$ and $\b\tau\tau$ channels. We then multiply the SS sample yields by this ratio. This method is used in the inclusive $\tau\tau$ channel but it suffers from large statistical uncertainties of the $\b\tau\tau$ SS sample. 
Therefore, we develop an alternate method that uses a MJ-enriched control sample with identical requirements as applied to the signal samples but reversing the muon isolation criteria. In a MJ-dominated SS sample, obtained without any requirement on the number of jets ($N_{\text{jets}}$), the ratio of the probabilities for a muon of a MJ-event to appear isolated or not isolated, $\RMJ\equiv  \mathcal{P}(\mu_{\mathrm{iso}} | \text{MJ}) / \mathcal{P}(\mu_{\overline{\mathrm{iso}}} | \text{MJ})$, 
is measured as function of $\eta^{\tauh}$, $\pttau$, and leading-jet $\pt$ (if $N_{\text{jets}}>0$). The ratio \RMJ is then applied to the distributions of the non-isolated-muon sample, predicting the MJ background in the two signal samples. This method is used in the $\b\tau\tau$ study. 
In each analysis, the alternate method is used to determine the systematic uncertainty on the MJ-background normalisation.

The distributions of  $\mhat$ for the $\tau\tau$ study and two different $\Df$ discriminants for the $\b\tau\tau$ analysis are presented in Fig.~\ref{fig:ditau_vars}. The observed data, expected signal and background yields are given in Table~\ref{tab:ditau_yields} for the two di-tau event selections.

\subsection{\bbb final state}
In the \bbb analysis, at least three jets, each satisfying $p_T>15~\gev$, $|\eta|<2.5$ and $NN_b>0.775$, are required. The two leading jets must have $p_T > 25 $ GeV. To improve the signal sensitivity, the events are separated into two channels, containing exactly 3 or 4 jets. The data and signal yields are given in Table~\ref{tab:bbb_yields}.  In addition, a likelihood discriminant, $\mathcal{D}_{\bbb}$, based on six kinematic variables is  employed.  Two separate likelihoods, one for the mass region $90 \leq M_\mathrm{A} < 140~\gev$ and the other for $140 \leq M_\mathrm{A} < 300~\gev$, are used. The dominant heavy flavor multijet backgrounds are estimated using a data driven technique. 
The background in the triple b-tagged sample is estimated by applying a 2D-transformation in $M_{\b\bbar}$ and $\mathcal{D}_{\bbb}$, derived from the ratio of the number of MC events in the triple and double $b$-tagged samples, to the double $b-$tagged data sample. The method significantly reduces the sensitivity of the background model to the underlying kinematics of the simulated events and the modelling of the geometric acceptance of the detector. The appropriate composition of the simulated samples is determined by comparing the sum of the transverse momenta of the jets in each event in simulation and data for various $b$-tagging criteria.  The invariant mass distribution of the jet pairing with the highest $\mathcal{D}_{\bbb}$ value is used as the final discriminant. The distribution for the dominant 3-jet channel is shown in Fig.~\ref{fig:inputshbb}a. In Fig.~\ref{fig:inputshbb}b, good agreement is observed between the data and background model in a control sample selected using an inverted likelihood criterion $\mathcal{D}_{\bbb}<0.12$.
 \begin{table}[h]
\bc
\caption{ Observed data yield and expected signal yields in the \bbb channel. 
The signal yields are given for the scenario described in Table~\ref{tab:ditau_yields}.}
\label{tab:bbb_yields}
%\begin{ruledtabular}
\begin{tabular}{ l c c}
\hline\hline
                    & \bbb  & \bbb    \\
$N_\text{jets}$ & 3 & 4 \\
\hline
Data           & 15214          & 10417   \\
\hline   
Signal $m_\phi=100~\gev$ & 335 & 166\\
Signal $m_\phi=190~\gev$ &   70 &    36\\
\hline\hline
\end{tabular}
%\end{ruledtabular}
\ec
\end{table}
\begin{figure}[!th]
	\bc
	\includegraphics[width=0.85\linewidth]{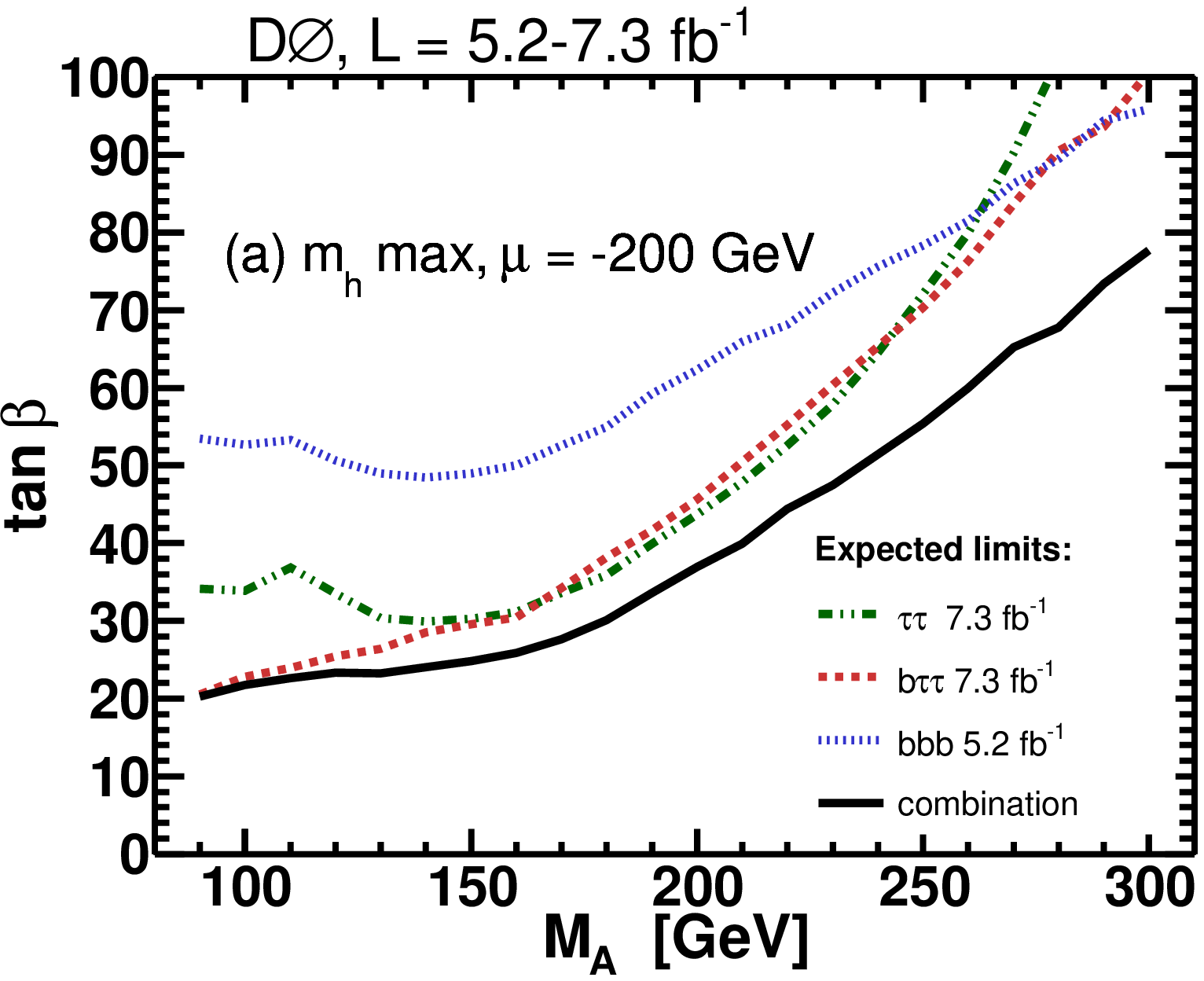}
	\includegraphics[width=0.85\linewidth]{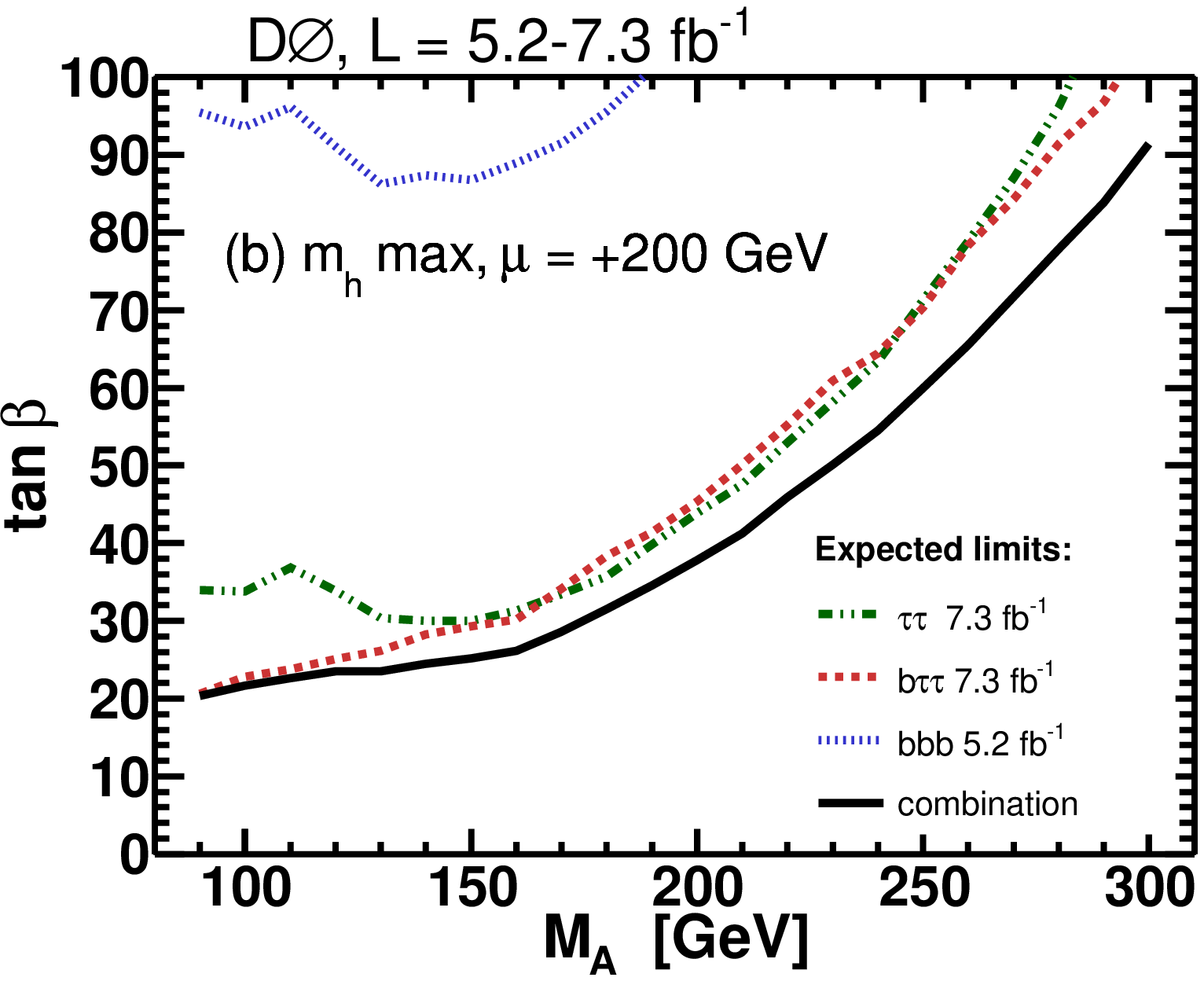}
	\caption{Comparison of the expected limits in the  $(\tanb,M_A)$ plane for the three channels separately, and their combination for the $m_h^{\text{max}}$ scenario with (a) $\mu<0$ and (b) $\mu>0$. }
	\label{fig:exp_sensitivity}
	\ec
\end{figure}
\subsection{Systematic uncertainties}

Depending on the source, we consider the effect of systematic uncertainties on the normalization and/or on the shape of the differential distributions of the final discriminants.

In the di-tau channels, the $Z$(+jets) background uncertainties are estimated using $\zmm$ data control samples, resulting in normalisation uncertainties of $3.2\%$ ($5\%$) for $Z$($+b$-tagged jets) boson production, an inclusive trigger efficiency uncertainty of  $3\%$ (common to all simulated backgrounds) and a shape-dependent uncertainty of $\sim$$1\%$ from the modeling of the $Z$ boson kinematics. 
The MJ-background uncertainty ranges from $10\%$ to $40\%$ on the $\b\tau\tau$ channel yields  while it is found to be shape dependent in the $\tau\tau$ channel (up to $100\%$ at high $M_{hat}$). 
For the remaining backgrounds and for signal, we consider uncertainties affecting the normalisation: luminosity ($6.1\%$), muon reconstruction efficiency ($2.9\%$), \tauh reconstruction efficiency [($4$--$10)\%$], single muon trigger efficiency ($1.3\%$), \ttbar ($11\%$) and diboson ($7\%$) production cross sections. Further sources of uncertainty affecting the shape of the final discriminant are considered: the jet energy scale ($10\%$) and the modeling of the $b$-tagging efficiency ($\sim$$4\%$) mostly affect the $\b\tau\tau$ signal modelling but are negligible in the $\tau\tau$ channel, while the \tauh energy scale ($\sim$$10\%$) only impacts significantly the $\tau\tau$ search for both $Z$ boson background and signal $M_{hat}$ distribution. 
With the exception of the \tauh reconstruction efficiency, \tauh energy scale and MJ estimation, which are evaluated for each \tauh type, these uncertainties are assumed to be $100\%$ correlated across both di-tau channels. 

In the \bbb channel, for the dominant MJ background, only systematic variations in the shape of the $M_{b\bbar}$  distribution are considered, as only the shape, and not the normalisation, is used to distinguish signal from background~\cite{cite:D0_bbb3}. The dominant sources arise from the measurement of the rate at which light partons fake a heavy flavor jet and the $b$-tagging efficiency.
For the signal model, the $b$-tagging efficiency (11-18\%), the luminosity ($6.1\%$) and the jet energy scale [$(2$--$10)\%$] dominate the experimental uncertainties.

Most of the experimental uncertainties are uncorrelated between the di-tau and the \bbb analyses with the exceptions of the $b$-quark efficiency, luminosity, and jet energy scale, which are assumed to be $100\%$ correlated. The theoretical uncertainties on the signal are other sources of correlated systematic uncertainty among all channels. They are dominated by parton density function uncertainties, renormalisation and factorisation scales. We assign an uncertainty of $15\%$ on the theoretical cross sections that is correlated across all processes.\\

\begin{figure}[tb]
	\bc
	\includegraphics[width=0.90\linewidth]{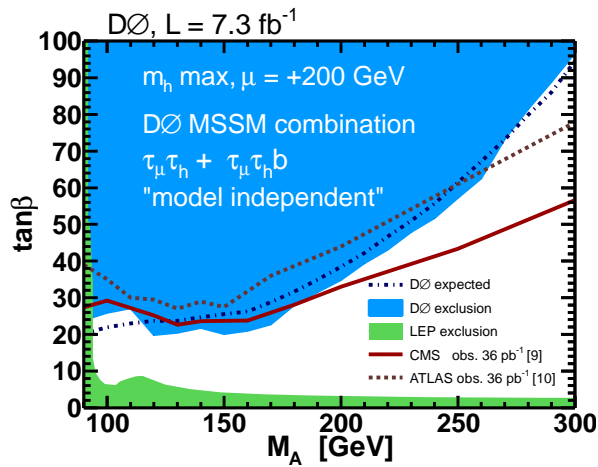}
	\caption{Constraints in the $(\tanb,M_A)$ plane from the di-tau combination in the $m_h^{\text{max}}$ scenario. These limits  very weakly depends on the other MSSM parameters.}
	\label{fig:CLfit2_tanBeta_tautauX}
	\ec
\end{figure}

\begin{figure*}[tbh]
	\bc
	\includegraphics[width=0.45\linewidth]{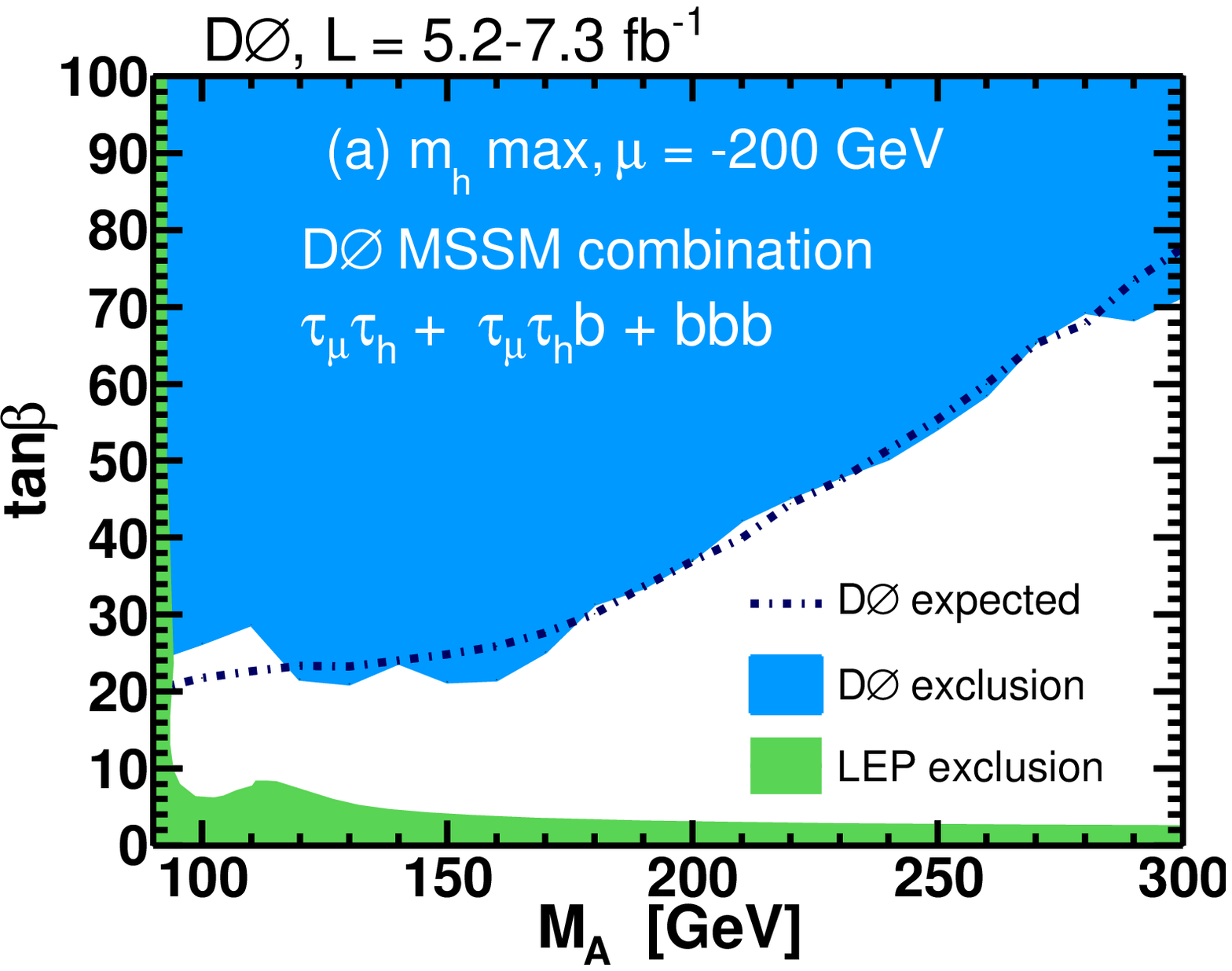}
	\includegraphics[width=0.45\linewidth]{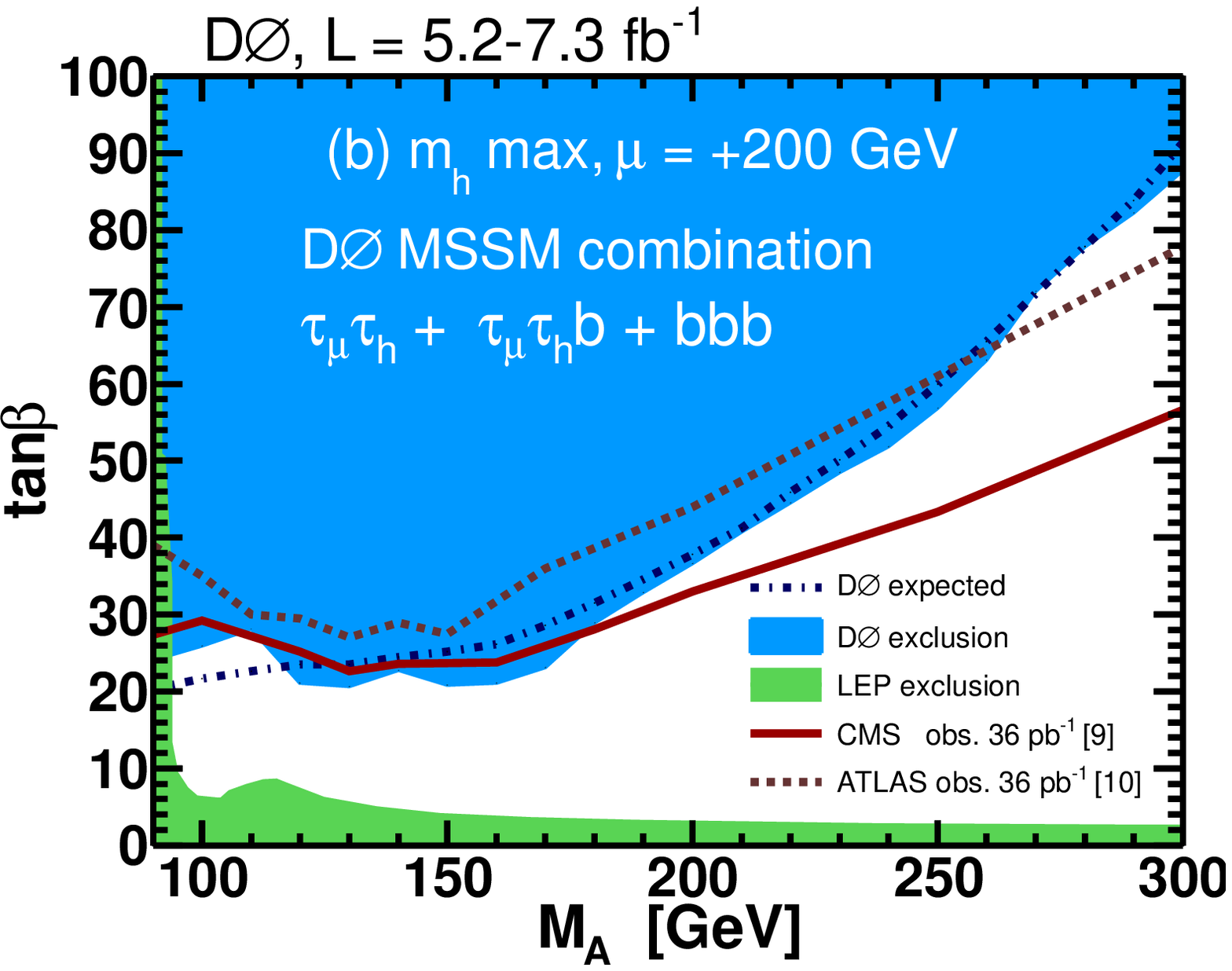}
	\includegraphics[width=0.45\linewidth]{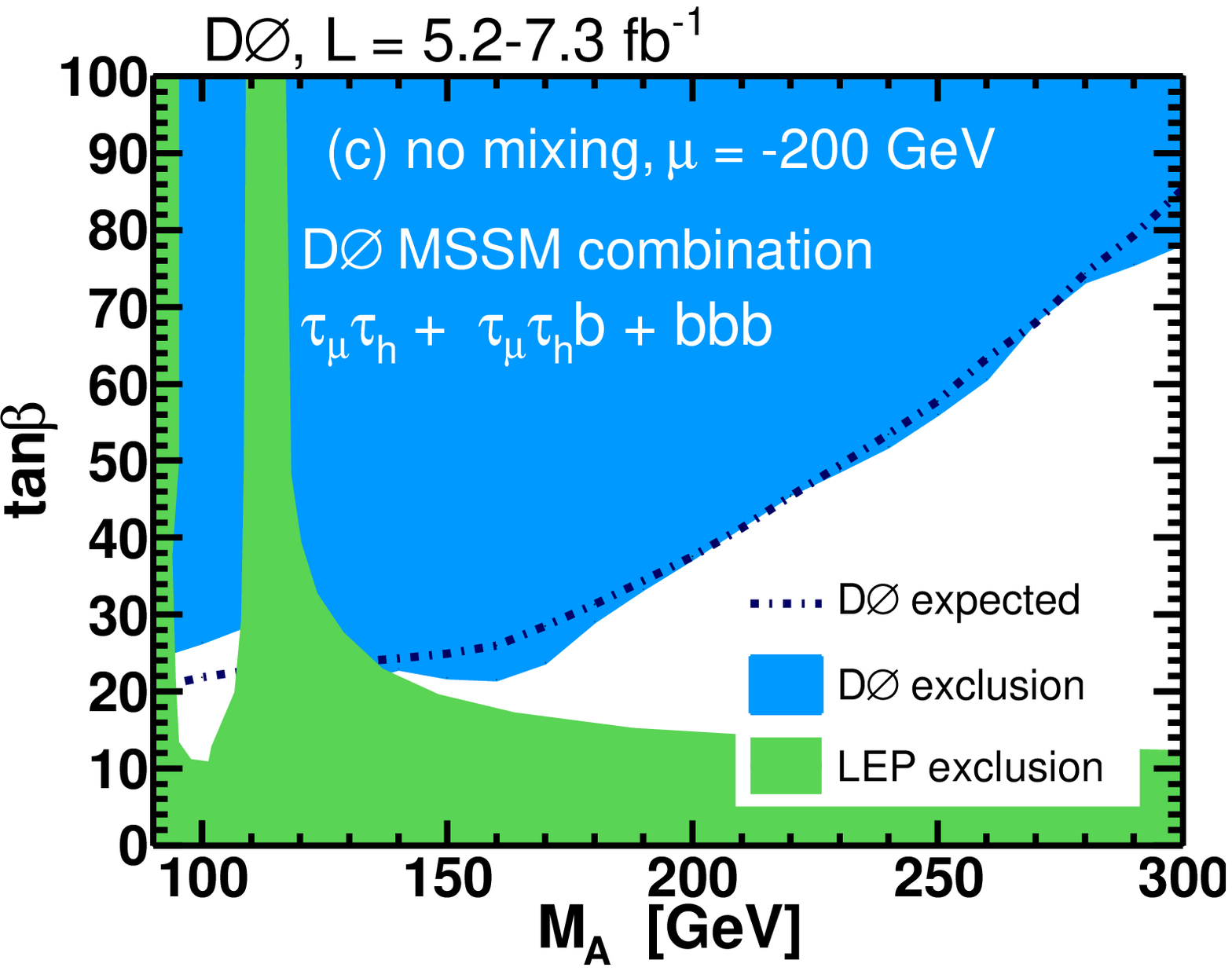}
	\includegraphics[width=0.45\linewidth]{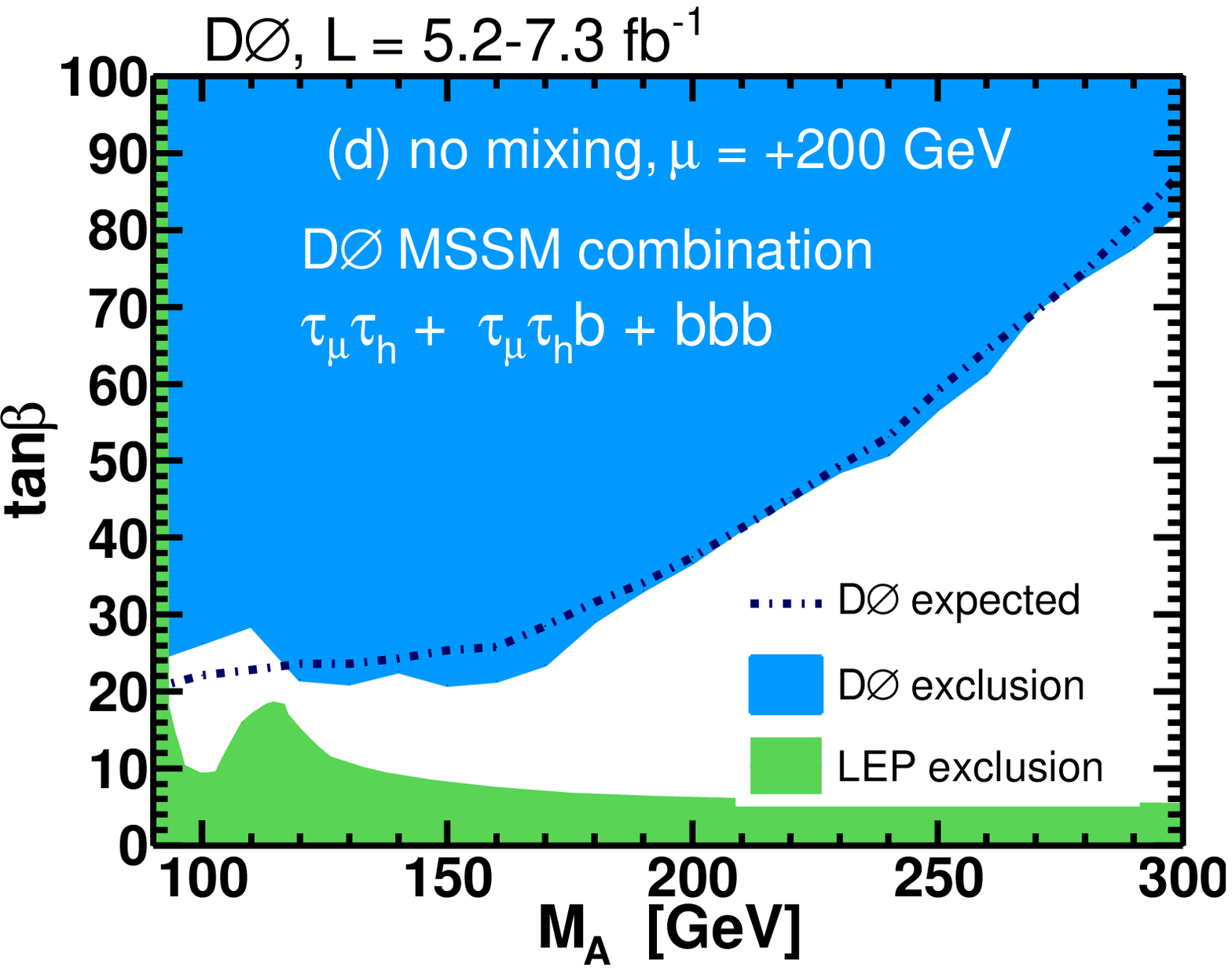}

	\caption{Constraints in the $(\tanb,M_A)$ plane for different MSSM scenarios from the combined Higgs bosons  searches. }
	\label{fig:CLfit2_tanBeta}
	\ec
\end{figure*}
\section{Results}

We combine the $\tau\tau$, $\b\tau\tau$ and $\bbb$ channels using the modified frequentist approach~\cite{cls}. The test statistic is a negative log-ratio of profiled likelihoods~\cite{collie}:
$$LLR = -2\ln\frac{p({\mathrm{data}}|H_1)}{p({\mathrm{data}}|H_0)},$$
where $H_1$ is the test (background + signal) hypothesis, $H_0$ is the null (background only) hypothesis and $p$ are the profile likelihoods based on Poisson probabilities for obtaining the observed number of events under each hypothesis. We define $CL_s$ by $CL_s \equiv CL_{s+b}/CL_{b}$, where $CL_{s+b}$ and $CL_{b}$ are the confidence levels for the test and null hypothesis respectively. We exclude signal yields with $CL_s < 0.05$. 

The $LLR$ quantity is computed from the $M_{hat}$ distribution for the $\tau\tau$ channel, the $\Df$ distributions for the $\b\tau\tau$ channel and the $M_{b\bbar}$ distribution for the \bbb channel. The NNLO SM cross sections  $\sigma_{\ggh}$ and $\sigma_{\bbh}$ are taken from~\cite{Harlander:2002wh,Anastasiou:2002yz,Ravindran:2003um,Catani:2003zt,Aglietti:2004nj,Actis:2008ug,Graudenz:1992pv,Spira:1995rr} and~\cite{Harlander:2003ai}, respectively, while the NLO SM cross section $\sigma_{\gbhb}$ is taken from {\sc mcfm}. The model-dependent MSSM to SM cross section ratios are computed with {\sc feynhiggs}~\cite{feynhiggs}.
To avoid double counting between the \bbh and \gbhb processes, we obtain the expected signal yield $N_{\tau\tau\text{+X}}^{exp}$ in the di-tau channels by
\begin{eqnarray*}
\frac{N_{\tau\tau\text{+X}}^{exp}}{ \mathcal{L} } &=& \mathcal{A}_{\ggh}\times \sigma_{\ggh}^\text{\ model} + \mathcal{A}_{\bbh}\times\left( \sigma_{\bbh}^\text{\ model} - \sigma_{\gbhb}^\text{\ model} \right) \\
			                             &+&     \mathcal{A}_{\gbhb}\times \sigma_{\gbhb}^\text{\ model},
\end{eqnarray*}
where the acceptances $\mathcal{A}$ are computed using the simulation and include the experimental efficiency. The two first terms of this equation refers to Higgs boson production without any  $b$ quark within the acceptance, while the third term is used for \gbhb production. There is no difference in the experimental acceptance for the \ggh and \bbh processes with no outgoing \b quark within the acceptance. Therefore, we set $\mathcal{A}_{\bbh} \equiv \mathcal{A}_{\ggh}$. The Higgs boson width, calculated with {\sc feynhiggs}, is also taken into account~\cite{cite:D0_bbb3}. \\

We test two MSSM benchmark scenarios~\cite{mssm_s}, no-mixing and $m_h^{\text{max}}$, and we vary the sign of the higgsino mass parameter, $\mu$. The expected sensitivities for two $m_h^{\text{max}}$ scenarios are shown on Fig.~\ref{fig:exp_sensitivity} for the three different searches and for their combination. At low $M_A$, the $\b\tau\tau$ channel dominates the sensitivity. For intermediate $M_A$, the $\tau\tau$ and $\b\tau\tau$ channels have similar sensitivities, while at high $M_A$, the $\bbb$ sensitivity becomes appreciable especially in $\mu<0$ scenarios. While the sensitivity in the $\tau\tau$+X channels are barely sensitive to other MSSM parameters than $M_A$ and $\tan\beta$, the \bbb signal yields is much more model dependent. Therefore we also provide a combination of the $\tau\tau$ and $\b\tau\tau$ searches only.
We do not observe any significant excess in data above the expected background fluctuations and we proceed to set limits. The limit from the $\tau\tau$+X combination is shown in Fig.~\ref{fig:CLfit2_tanBeta_tautauX} and the full combination limits in different MSSM scenarios are shown in Fig.~\ref{fig:CLfit2_tanBeta} .\\

In summary, we present  MSSM Higgs boson searches in three final states: $\tau\tau$, $\b\tau\tau$ and $\bbb$. These different searches are combined to set limits in the $(\tanb,M_A)$ plane in four different MSSM scenarios. Furthermore, we combine the  $\tau\tau$ and $\b\tau\tau$ channels to obtain MSSM-scenario independent limits. We exclude a substantial region of the MSSM parameter space, especially for $M_A<180~\gev$ where we exclude $\tanb>20-30$. These are the tightest constraints from the Tevatron on the production of neutral Higgs bosons in the MSSM and are comparable to the published LHC limits~\cite{cite:CMS_tautau,cite:ATLAS_tautau}, especially at low $M_A$. 

\section*{Acknowledgements}
% acknowledgement.tex                             6 April 2011
%
We thank the staffs at Fermilab and collaborating institutions,
and acknowledge support from the
DOE and NSF (USA);
CEA and CNRS/IN2P3 (France);
FASI, Rosatom and RFBR (Russia);
CNPq, FAPERJ, FAPESP and FUNDUNESP (Brazil);
DAE and DST (India);
Colciencias (Colombia);
CONACyT (Mexico);
KRF and KOSEF (Korea);
CONICET and UBACyT (Argentina);
FOM (The Netherlands);
STFC and the Royal Society (United Kingdom);
MSMT and GACR (Czech Republic);
CRC Program and NSERC (Canada);
BMBF and DFG (Germany);
SFI (Ireland);
The Swedish Research Council (Sweden);
and
CAS and CNSF (China).
%


%\bibliographystyle{model2-names}

\begin{thebibliography}{99}
\bibitem{pseudorapidity}  
  The pseudorapidity $\eta$ is defined relative to the center of the detector as $\eta=-\ln[\tan(\theta/2)]$ where $\theta$ is the polar angle with respect to the proton beam direction.
\end{thebibliography}

%% not want to use model2-names.bst.

%\bibliographystyle{unsrt}

\end{document}